\documentclass[a4paper]{revtex4}
\usepackage[top=1in, bottom=1in, left=1in, right=1in]{geometry}
\usepackage{amsmath}
\usepackage{amssymb}
\usepackage{amsfonts}
\usepackage{graphicx}
\usepackage{dcolumn}
\usepackage[colorlinks,linkcolor=blue,citecolor=blue,urlcolor=blue]{hyperref}

\begin{document}

\title{Noether symmetry analysis of anisotropic universe in $f(T,B)$ gravity}
\author{Yusuf Kucukakca \thanks{E-mail address:ykucukakca@akdeniz.edu.tr}}
\affiliation{Department of Physics, Faculty of Science, Akdeniz University, 07058 Antalya, Turkey}
\begin{abstract}
The present work is devoted to investigate the Noether symmetries of the locally rotationally symmetric Bianchi type I space time in $f(T,B)$ gravity theory which depends on the torsion scalar $T$ and the boundary term $B$. In this theory, we consider some particular models and investigate their Noether symmetry generators. Besides, we get exact cosmological solutions of the considering models including the matter dominant universe using the Noether symmetry technique. The obtained results are coincide with the accelerated expansion behaviour of the universe.
\end{abstract}


\maketitle
\section{Introduction}
Astrophysical observations in recent years have indicated that our universe is expanding at an accelerated phase. Various cosmological scenarios have been proposed to clarify this interesting behaviour of the universe \cite{riess99,Perll,Sperl,netterfield02}. In this context, two types of categories have been considered in the literature. The first category is to introduce in the framework of General Theory of Relativity (GR) theory an exotic liquid, called dark energy, has a repulsive gravitational feature because it creates a negative pressure. It is believed that the late-time accelerating expansion of the universe may be due to the existence of dark energy. Although the underlying physics of the dark energy is still unclear, One of the remarkable nominees is the cosmological constant which yields a negative pressure with the equation of state (EoS) parameter, $\omega=-1$. However, because of the fact that the cosmological constant causes some problems such as extreme fine-tuning and coincidence problem has gradually lost its popularity \cite{carroll01}. In order to overcome these problems, it has been proposed many dark energy models which are some kinds of the scalar field as quintessence  \cite{Rat}, quintom \cite{Guo}, phantom energy \cite{Cad}, fermion \cite{Kremer,kuc1,kuc2} or tachyon field \cite{SenA}. The second category is mainly based on the modifications of GR as a purely geometric effect. These modified theories may be considered as the most popular candidates to reveal the mysterious nature of dark energy. The most important one of these theories is $f(R)$ gravity that is constructed by inserting an arbitrary function of the curvature scalar to the Einstein-Hilbert action. In recent times, it has been put forward to several forms of $f(R)$ gravity, and discussed in many fields including the early and late-time cosmic acceleration, solar system test, black hole solution \cite{noji,Capo1,olmo1,cruz}.

Teleparallel gravity (TG) is equivalent to GR such that its modified form is alternative to explain the cosmic acceleration providing a gravitational alternative to dark energy. This theory, called $f(T)$ gravity theory is constructed by inserting an arbitrary function of the torsion scalar to action of TG \cite{Beng,Lind,Myr}. In this formalism of gravity, one could use the Weitzenböck connection that has torsion but not curvature, instead of utilizing curvature constructed by the Levi-Civita connection in GR. The dynamical variables in TG are tetrad fields. An important advantage of this theory is that it has the second-order field equations and therefore it is easy to deal with when compared to $f(R)$ gravity theory with the fourth-order field equations. On the other hand, Li et al. demonstrated that $f(T)$ gravity theory and its field equations are not invariant under local Lorentz transformations \cite{lii}. Recently, by this motivation, in the framework of teleparallel gravity it has been formulated a new modified gravitational theory named as $f(T,B)$ gravity that is reduced to both $f(T)$ and $f(R)$ gravity by the special selection of its form \cite{baha1}. Lagrangian density of $f(T,B)$ theory depends on both the torsion scalar T and the boundary term B. In Ref. \cite{baha2}, the authors have discussed different cosmological features for this theory such as use of the reconstruction technique, examination of the validity of the laws of thermodynamics. Also, some cosmological solutions have been examined by using Noether symmetry approach in spatially FRW metric \cite{baha3}.

The most suitable model for identifying the large-scale structure of the universe may be thought of as FRW space-time which has a spatially homogeneous and isotropic nature. On the other hand, there are some indications in the CMB temperature anisotropy studies that may break isotropic nature of the universe, leading to some interesting anomalies \cite{copi}. Therefore, it is important to explore the Bianchi space times giving an information about the anisotropy in the early and late time universe on the current observations \cite{uzan}. We also note that these models are the more generalized form of FRW universe. Akarsu and Kilinc \cite{akarsu} have examined the anisotropic dark energy model in the LRS Bianchi type I cosmological analysis. In Ref. \cite{shaarif}, the authors discussed some anisotropic solutions in the context of $f(R)$ gravity. Bianchi cosmological models have been studied both in GR and in modified theories of gravity to understand the dynamics of the universe \cite{kumar,koi,adh,sh1,akca1}.

Symmetries play an important role in finding some exact solutions to dynamical systems. In particular, Noether symmetry that can be related to differential equations having a Lagrangian is a useful approach that leads to the existence of conserved quantities. In addition, this method is very useful for determining unknown functions which are exist in the Lagrangian. Up to now, this approach has been extensively studied in cosmological models such as scalar-tensor theories \cite{capp2,capp3,akca2,pal1,bel}, telerapallel dark energy model \cite{akca3,taj}, models of fermionic field \cite{Kremer,kuc1,kuc2}, $f(R)$ and $f(T)$ theories  \cite{capp4,Vak1,akca4,akca5,wei2,pal2}. Furthermore, the technique has also been performed in different Bianchi space times \cite{sha3,sk1,sk2}, Gauss-Bonnet gravity \cite{capp7,sha5,sk3} and others see \cite{taj1,akca6,mome}. On the other hand, Sharif and Nawazish \cite{sarf} investigated the existence of Noether symmetry for the anisotropic models in $f(R)$ theory. Aghamohammadi \cite{ali} found exact solution of the anisotropic space time with $f(T)$ power-law model using the Noether symmetry approach. Recently, Bahamonde and Capozziello \cite{baha3} explored some cosmological solutions by fixing the forms of $f(T,B)$ gravity with the presence of Noether symmetry for the FRW space-time. In this work, following the calculations performed in Ref. \cite{baha3}, we search the Noether symmetries for the LRS Bianchi type I model in the $f(T,B)$ theory. We also examine some important cosmological parameters to determine how evolution of the universe evolved over time.

The outline of this work is as follows: In Section \ref{sc2}, we give the basic formalism of the teleparallel formulation of general relativity and its modified theories. We derive the modified field equations of $f(T,B)$ gravity for the LRS Bianchi type I model in Section \ref{sc3}. In Section \ref{sc4}, we investigate the Noether symmetries of the model and analyse the cosmological solutions. Finally, in Section \ref{sc5}, we give the basic results of this work.
\section{$f(T,B)$ gravity}\label{sc2}
We now shortly review the teleparallel formulation of GR and its modifications. In the TG theories the fundamental dynamical objects are the tetrad fields (vierbeins) $e_{\mu}^{a}$ its inverse tetrad fields are $E_{a}^{\mu}$. The tetrad and the inverse tetrad fields satisfy the following orthogonality conditions,
\begin{eqnarray}\label{ort}
E_{m}^{mu}e_{\mu}^{n} =\delta_{m}^{n},\qquad E_{m}^{\nu}e_{\mu}^{m} =\delta_{\mu}^{\nu}.
\end{eqnarray}
The metric tensor $g_{\mu\nu}$ can be generated from the tetrad fields as
\begin{equation} \label{metr}
g_{\mu\nu}=\eta_{ab}e_{\mu}^{a}e_{\nu}^{b},
\end{equation}
here $\eta_{ab}$ is the Minkowski metric with the signature $-2$. As it is well known, GR is based on the symmetric Levi-Civita connection is used to construct the covariant derivative. In contrast to the GR, teleparallel gravity is utilized the anti-symmetric Weitzenböck connection defined as \cite{malof},
\begin{equation} \label{weit}
W_{\mu}{}^{a}{}_{\nu}=\partial_{\mu}e^{a}{}_{\nu},
\end{equation}
which yields zero curvature but nonzero torsion. The torsion tensor is the antisymmetric part of this connection as follows
\begin{equation}\label{torten}
T^{a}\,_{\mu\nu}=W_{\mu}{}^{a}{}_{\nu}-W_{\nu}{}^{a}{}_{\mu}=\partial_{\mu}e_{\nu}^{a}-\partial_{\nu}e_{\mu}^{a}.
\end{equation}
The Weitzenböck connection of TG can be expressed in term of the usual Levi-Civita connection, which we denote by ${}^0\Gamma$ of GR as
\begin{equation}
W_{\lambda}{}^{\mu}{}_{\rho}={}^0\Gamma_{\lambda \rho}^{\mu}+K_{\lambda}{}^{\mu}{}_{\rho},
\end{equation}
where $K$ is called the contorsion tensor which is defined by the torsion tensor
\begin{equation}
2K_{\mu}{}^{\lambda}{}_{\nu}=T^{\lambda}{}_{\mu\nu}-T_{\nu\mu}{}^{\lambda}+T_{\mu}{}^{\lambda}{}_{\nu}.
\end{equation}
One also defines the following tensor
\begin{equation}\label{defii}
S_{\sigma}^{\mu \nu}=\frac{1}{2}\left(K^{\mu\nu}{}_\sigma-\delta_{\sigma}^{\mu}T^{\nu}+\delta_{\sigma}^{\nu}T^{\mu}\right),
\end{equation}
here $T^{\mu}$ is called as the torsion vector, is obtained by the contraction of the torsion tensor. The combination of equation (\ref{defii}) with the torsion tensor (\ref{torten}) leads to the torsion scalar
\begin{eqnarray}
T=T^{\alpha}{}_{\mu\nu}S_{\alpha}{}^{\mu\nu},
\end{eqnarray}
then the action of TG reads
\begin{eqnarray}
\textbf{S}=\frac{1}{\kappa^2}\int{d^4 x e T}+\textbf{$S_{m}$}, \label{tpa},
\end{eqnarray}
where e is the volume element of the metric tensor that is equal to $\sqrt{-g}$ and \textbf{$S_m$} is the action of the standard matter content. Using the definitions given above, one can easily achieve the relation among the Ricci scalar related to the Levi-Civita connection and the torsion scalar \cite{baha1},
\begin{equation}
 R = - T + \frac{2}{e}\partial_\mu (e T^\mu)=-T+B \, \label{ricciT},
\end{equation}
where $B=\frac{2}{e}\partial_\mu (e T^\mu)$ is a boundary term. This relationship given by Eq. (\ref{ricciT}) tells us that the action of the TG (\ref{tpa}) is dynamically equivalent to the standard action of GR, since they only differ by a total derivative.

One of the most popular generalizations of the teleparallel gravity is $f(T)$ gravity. The action integral for this theory is given by \cite{Beng}
\begin{eqnarray}
\textbf{S}=\frac{1}{\kappa^2}\int{d^4 x e f(T)}+\textbf{$S_{m}$}, \label{gtpa},
\end{eqnarray}
where $f(T)$ is a function of $T$. It is clear that $f(T)$ is a linear function of $T$ then the action (\ref{tpa}) is recovered. The gravitational field equations are derived by taking the variation according to the tetrad field of the action (\ref{gtpa}). The resulting field equations are a second order because the torsion scalar consists in the first derivatives of the tetrad fields. In recent years, a new and interesting modified teleparallel theory of gravity has been proposed by Bahamonde et al. to combine these two theories. The new action has the following form \cite{baha1}
\begin{eqnarray}
\textbf{S}=\frac{1}{\kappa^2}\int{d^4 x e f(T,B)}+\textbf{$S_{m}$}, \label{ngtpa},
\end{eqnarray}
where $f$ depends on $T$ and $B$. From the action (\ref{ngtpa}), one can show that the $f(T)$ and $f(R)$ gravity can be obtained by selecting $f(T,B)=f(T)$ and $f(T,B)=f(-T+B)=f(R)$, respectively. The gravitational field equations for the theory given by Eq. (\ref{ngtpa}) are as follows \cite{baha1}
\begin{eqnarray}
2e\delta_{\nu}^{\lambda}\nabla^{\mu}\nabla_{\mu} f_{B}-2e\nabla^{\lambda}\nabla_{\nu}f_{B}+
e B f_{B}\delta_{\nu}^{\lambda} \nonumber
\\
+4e\Big[(\partial_{\mu}f_{B})+(\partial_{\mu}f_{T})\Big]S_{\nu}{}^{\mu\lambda}
+4e^{a}_{\nu}\partial_{\mu}(e S_{a}{}^{\mu\lambda})f_{T} \nonumber
\\
-4 e f_{T}T^{\sigma}{}_{\mu \nu}S_{\sigma}{}^{\lambda\mu} -
e f \delta_{\nu}^{\lambda} = 16\pi e\Theta_{\nu}^{\lambda} \,,
\label{fieldeq}
\end{eqnarray}
where $\Theta_{\nu}^{\lambda}=e^{a}_{\nu}\Theta_{a}^{\lambda}$ is the standard energy momentum tensor, $\nabla_{\nu}$ stands for the covariant derivative with respect to the Levi-Civita connection and $f_{T}=\partial{f}/\partial{T}$, $f_{B}=\partial{f}/\partial{B}$. In the next section, we will focus on the anisotropic Bianchi type I cosmological model for the above mentioned the $f(T,B)$ theories.

\section{Anisotropic $f(T,B)$ Cosmology}\label{sc3}
In the present work, we explore the cosmological consequences of $f(T,B)$ theory. Especially, we deal with $f(T,B)$ anisotropic cosmology in spatially homogenous Bianchi type I space-time such that LRS line element is given by
\begin{equation}\label{bianchi}
ds^2=dt^2-X(t)^2dx^2-Y(t)^2\left(dy^2+dz^2\right)
\end{equation}
where directional scale factors $X$ and $Y$ are functions of time $t$. The field equations of $f(T,B)$ cosmology are obtained either by the help of the Eqs. (\ref{fieldeq}) or by using a point-like Lagrangian associated with the action (\ref{ngtpa}). Using the Eqs. (\ref{metr}) and (\ref{bianchi}), we find the diagonal tetrad components as follows
\begin{equation}\label{stt}
e_{\mu}^{a}=diag(1,X,Y,Y).
\end{equation}
For this tetrad component, the torsion scalar and boundary term can be calculated in their respective form as follows
\begin{eqnarray}\label{toba}
T=-2\left(\frac{2\dot{X}\dot{Y}}{XY}+\frac{\dot{Y}^2}{Y^2}\right),\nonumber
\\
B=-2\left(\frac{\ddot{X}}{X}+\frac{2\ddot{Y}}{Y}+\frac{4\dot{X}\dot{Y}}{XY}+\frac{2\dot{Y}^2}{Y^2}\right)
\end{eqnarray}
here the dot represents derivatives with respect to $t$. One can obtain the point-like Lagrangian related to the action (\ref{ngtpa}) if one uses the Lagrange multiplier approach to set $T$ and $B$ as a constraint of the dynamics. Therefore, inserting Eqs. (\ref{stt}) and (\ref{toba}) into the action (\ref{ngtpa}), we write again the action (\ref{ngtpa}) in physical units as follows
\begin{eqnarray}
S =\int dt XY^2\Big[f-\lambda_{1}\left[T+2\left(2\frac{\dot{X}\dot{Y}}{XY}+\frac{\dot{Y}^2}{Y^2}\right)\right]-
\lambda_{2}\left[B+2\left(\frac{\ddot{X}}{X}+2\frac{\ddot{Y}}{Y}+4\frac{\dot{X}\dot{Y}}{XY}+2\frac{\dot{Y}^2}{Y^2}\right)\right]+L_{m}\Big] \label{ac1},
\end{eqnarray}
where $\lambda_{1}$, $\lambda_{2}$ are the Lagrange multipliers and $L_{m}$ is the standard matter Lagrangian. We note that since there is no single definition of the matter Lagrangian, we can choose it as  $L_{m}=-\rho_{m}=-\rho_{m0}(XY^2)^{-1}$ which corresponds matter dominant universe \cite{akca5}. The variation of the action (\ref{ac1}) with respect to $T$ and $B$ lead to $\lambda_{1}=XY^2 f_{T}$ and $\lambda_{2}=XY^2 f_{B}$. After some calculations, we obtain the point-like Lagrangian as follows,
\begin{eqnarray}
L = XY^2\left(f-Tf_{T}-Bf_{B}\right)-2f_{T}(2Y\dot{X}+X\dot{Y})\dot{Y}+2Y(Y\dot{X}+2X\dot{Y})(f_{BB}\dot{B}+f_{TB}\dot{T})-\rho_{m0}\label{lag1}.
\end{eqnarray}
It is well known that the basic properties of a dynamical system can be determined by the Euler-Lagrange equation, given by
\begin{eqnarray}
\frac{\partial{L}}{\partial{q_{i}}}-\frac{d}{dt}\frac{\partial{L}}{\partial{\dot{q_{i}}}}=0 \label{eul},
\end{eqnarray}
where $q_{i}$ and $\dot{q_{i}}$ are generalized coordinates and velocities of the configuration space. The configuration space of the Lagrangian (\ref{lag1})  is $\mathcal{TQ}=(X,Y,T,B)$, its the tangent space is given by$\mathcal{T Q}=(X,Y,T,B,\dot X,\dot Y,\dot T,\dot B)$. Inserting the Lagrangian (\ref{lag1}) into the Euler-Lagrange equation for the variables $X$ and $Y$, we obtain
\begin{eqnarray}
f_{T}\left(\frac{2\ddot{Y}}{Y}+\frac{\dot{Y}^2}{Y^2}\right)+\frac{2\dot{Y}}{Y}\dot{f_{T}}-\ddot{f_{B}}+\frac{1}{2}\left(f-Tf_{T}-Bf_{B}\right)=0 \label{acc1},
\end{eqnarray}
\begin{eqnarray}
f_{T}\left(\frac{\ddot{X}}{X}+\frac{\ddot{Y}}{Y}+\frac{\dot{X}\dot{Y}}{XY}\right)+\left(\frac{\dot{X}}{X}+\frac{\dot{Y}}{Y}\right)\dot{f_{T}}-\ddot{f_{B}}+\frac{1}{2}\left(f-Tf_{T}-Bf_{B}\right)=0 \label{acc2}.
\end{eqnarray}
The modified Friedmann equation for $f(T,B)$ cosmology is obtained by imposing that the Hamiltonian related to the Lagrangian (\ref{lag1}) vanishes, i.e.
\begin{eqnarray}
\sum_{i}\frac{\partial{L}}{\partial{\dot{q_{i}}}}\dot{q_{i}}-L=0\qquad \qquad \qquad \qquad \qquad \qquad \qquad \qquad \qquad \nonumber \\
\Rightarrow f_{T}\left(\frac{2\dot{X}\dot{Y}}{XY}+\frac{\dot{Y}^2}{Y^2}\right)-\left(\frac{\dot{X}}{X}+\frac{2\dot{Y}}{Y}\right)\dot{f_{B}}+
\frac{1}{2}\left(f-Tf_{T}-Bf_{B}-\rho_{m}\right)=0\label{fri}.
\end{eqnarray}
We now consider the relation between the scale factors as $X=Y^m$; $m\neq 0,1$ where $m$ measures the deviation from the isotropy. When $m=1$, the universe is isotropic otherwise it is anisotropic. This physical condition comes from the assumption that the ratio of shear scalar to expansion scalar is constant. The physical importance of this condition by considering perfect fluid having barotropic equation of state is discussed by Collins \cite{coll}. Several researchers have also used this relation to obtain the cosmological solutions to the field equations \cite{y1,y2,y3}. Thus, we can rewrite the Lagrangian (\ref{lag1}) as follows
\begin{eqnarray}
L=Y^{m+2}\left(f-Tf_{T}-Bf_{B}\right)-2(2m+1)f_{T}Y^m\dot{Y}^2+2(m+2)Y^{m+1}\dot{Y}\left(f_{BB}\dot{B}+f_{TB}\dot{T}\right)-\rho_{m0}\label{lag2},
\end{eqnarray}
which depends on $Y$, $T$ and $B$. For this Lagrangian, the field equations reduce to following equations
\begin{eqnarray}
(2m+1)f_{T}\left(\frac{2\ddot{Y}}{Y}+m\frac{\dot{Y}^2}{Y^2}\right)+2(2m+1)\frac{\dot{Y}}{Y}\dot{f_{T}}-(m+2)\ddot{f_{B}}+\frac{m+2}{2}\left(f-Tf_{T}-Bf_{B}\right)=0 \label{neww1},
\end{eqnarray}
\begin{eqnarray} (2m+1)f_{T}\frac{\dot{Y}^2}{Y^2}-(m+2)\frac{\dot{Y}}{Y}\dot{f_{B}}+\frac{1}{2}\left(f-Tf_{T}-Bf_{B}-\rho_{m}\right)=0\label{neww2}.
\end{eqnarray}
Since these equations are non-linear differential equations, their solutions are very difficult. In order to find cosmological solutions to these equations, we also need to choose the form of the unknown function $f(T,B)$. In the next section, we utilize the Noether symmetry approach to determine the form of $f(T,B)$.

\section{Noether Symmetry Approach and Cosmological Solutions}\label{sc4}
This section deals with the Noether symmetry technique for the Lagrangian given by (\ref{lag2}). This technique is very useful for obtaining conserved quantities relevant to the dynamical system as well as for choosing the form of the unknown functions in the theory. Following Ref. \cite{capp2}, we define a vector field for the Lagrangian (\ref{lag2})
\begin{equation}\label{vec}
\textbf{X}=\alpha\frac{\partial }{\partial Y}+\beta\frac{\partial}{\partial T}+\gamma\frac{\partial}{\partial B}+
\dot \alpha\frac{\partial }{\partial \dot Y}+\dot \beta\frac{\partial }{\partial \dot T}+\dot \gamma\frac{\partial }{\partial \dot B},
\end{equation}
where $\alpha$, $\beta$ and $\gamma$ depend on the generalized coordinates $Y$, $T$ and $B$. The Noether theorem tells us that the Lie derivative of any Lagrangian along a vector field is zero i.e.
\begin{equation}\label{nc}
\mathcal{L}_\textbf{X}L=0,
\end{equation}
If this condition satisfy then \textbf{X} is a symmetry and it will be generated the following constant of motion (conserved quantity, first integral)
\begin{equation}\label{cm}
I_{0}=\alpha\frac{\partial L}{\partial \dot Y}+\beta\frac{\partial L}{\partial \dot T}+ \gamma\frac{\partial L}{\partial \dot B}.
\end{equation}
Hence implementing the Noether symmetry condition (\ref{nc}) for the Lagrangian (\ref{lag2}), we find the system of partial differential equations as
\begin{eqnarray}\label{nn1}
(2m+1)f_{T}\left(m\alpha+2Y\frac{\partial\alpha}{\partial Y}\right)+f_{TB}Y\left((2m+1)\gamma-(m+2)Y\frac{\partial\beta}{\partial Y}\right) \nonumber \\
+(2m+1)Y f_{TT}\beta-(m+2)Y^2f_{BB}\frac{\partial\gamma}{\partial Y}=0,
\end{eqnarray}
\begin{equation}\label{nn2}
f_{TB}\frac{\partial\alpha}{\partial T}=0, \qquad f_{BB}\frac{\partial\alpha}{\partial B}=0,
\end{equation}
\begin{eqnarray}\label{nn3}
(m+2)f_{TB}\left((m+1)\alpha+Y\frac{\partial\alpha}{\partial Y}+Y\frac{\partial\beta}{\partial T}\right)-2(2m+1)f_{T}\frac{\partial\alpha}{\partial T}\nonumber \\ +(m+2)Y f_{BB}\frac{\partial\gamma}{\partial T}+(m+2)Y\left(f_{TTB}\beta+f_{TBB}\gamma\right)=0,
\end{eqnarray}
\begin{eqnarray}\label{nn4}
(m+2)f_{BB}\left((m+1)\alpha+Y\frac{\partial\alpha}{\partial Y}+Y\frac{\partial\gamma}{\partial B}\right)-2(2m+1)f_{T}\frac{\partial\alpha}{\partial B}\nonumber \\ +(m+2)Y f_{TB}\frac{\partial\beta}{\partial B}+(m+2)Y\left(f_{TBB}\beta+f_{BBB}\gamma\right)=0,
\end{eqnarray}
\begin{eqnarray}\label{nn5}
(m+2)\left(f-Tf_{T}-Bf_{B}\right)\alpha-Y\left(Tf_{TT}+Bf_{TB}\right)\beta-Y\left(Tf_{TB}+Bf_{BB}\right)\gamma=0.
\end{eqnarray}
There are two different ways to solve the Noether symmetry equations given by Eqs. (\ref{nn1})-(\ref{nn5}): the first is to choose particular shape of $f(T,B)$ and then to find the components of the vector field accordingly. The second method is to solve equations directly and find the unknown functions. From a physical perspective, the first method is more preferable because it permits studying credible models. So we choose the second method to study the anisotropic $f(T,B)$ models.
\subsection{Case 1: $f(T,B)=f(T)$} \label{sec:4.1}
The first important model is the $f(T)$ gravity. In this case, Lagrangian (\ref{lag2}) does not include the boundary term B. From the Noether symmetry equations (\ref{nn1})-(\ref{nn5}), we can easily find the following solution for $\alpha$, $\beta$ and $f(T)$
\begin{eqnarray}\label{vec1}
\alpha=\alpha_{0}Y^{1-\frac{m+2}{2n}}, \quad \beta=-\frac{\alpha_{0}(m+2)}{n}Y^{-\frac{m+2}{2n}}T,
\end{eqnarray}
\begin{eqnarray}\label{form1}
f(T)=T_{0}T^n,
\end{eqnarray}
where $n$, $\alpha_{0}$ and $T_{0}$ are an integration constants.From the Eq. (\ref{cm}), the first integral associated with the Noether symmetry corresponding to this solution has the form
\begin{eqnarray}\label{p1}
Y^{\frac{m+2-2n}{2n}}\dot{Y}=k_{0},
\end{eqnarray}
where we define
\begin{eqnarray}\label{0l}
k_{0}=\left(\frac{I_{0}}{-4(-2)^{n-1}\alpha_{0}T_{0}n(2m+1)^{n}}\right)^{\frac{1}{2n-1}}\nonumber.
\end{eqnarray}
The general solution of the equation (\ref{p1}) is
\begin{eqnarray}\label{s1}
Y(t)=\left[\frac{k_{0}(m+2)}{2n}t+c_{1}\right]^{\frac{2n}{m+2}},
\end{eqnarray}
where $c_{1}$ is an integration constant and $m\neq -2$. From the condition $X=Y^m$, we obtain the scale factor along x-direction as following
\begin{eqnarray}\label{s2}
X(t)=\left[\frac{k_{0}(m+2)}{2n}t+c_{1}\right]^{\frac{2mn}{m+2}}.
\end{eqnarray}
Consequently, we have a power-law form for the scale factors. Such models suitable for Noether symmetry have been studied extensively in the literature for both isotropic  \cite{wei2} and anisotropic  \cite{ali} space time. For $m=-2$, from Eq. (\ref{p1}) and using the definition of the average factor, we obtain $a(t)=a_{0}e^{k_{0}t}$ which is a de Sitter solution.
\subsection{Case 2: $f(T,B)=b_{0}B^k+t_{0}T^n$} \label{sec:4.2}
Second, we assume that $f(T,B)=b_{0}B^k+t_{0}T^n$ where $b_{0}$, $t_{0}$, $k$ and $n$ are the arbitrary constants. Substituting this form of $f(T,B)$ in the Noether symmetry equations (\ref{nn1})-(\ref{nn5}), a trivial solution is obtained by $\alpha=\beta=\gamma=0$ for $k\neq 1$ which means that there is no Noether symmety. For $k=1$, we have $f(T,B)=b_{0}B+t_{0}T^n$ which is the same as the previous case. At this point we can note that if the function $f(T,B)$ is linear with respect to $B$, then there is no change in the field equations.	
\subsection{Case 3: $f(T,B)=b_{0}B^kT^n$} \label{sec:4.3}
\begin{figure}
\resizebox{0.45\textwidth}{!}{\includegraphics{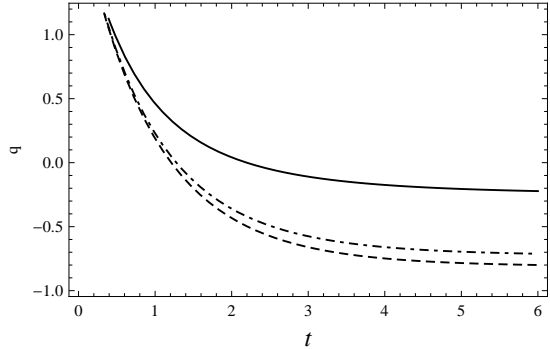}}
\caption{Plots of the deceleration parameter against $t$ for the different value of $k$. We set $k=3$ (solid line), $k=10$ (dashed line), $k=15$ (dot dashed line) and  $m=1.00672$, $z_{1}=\beta_{0}=-1$ and $z_{2}=1$.} \label{fig:5}
\end{figure}

\begin{figure}
\resizebox{0.45\textwidth}{!}{\includegraphics{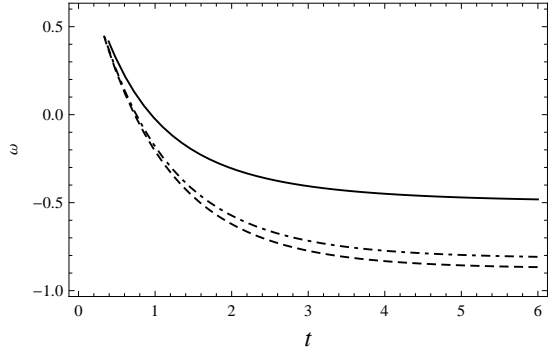}}
\caption{Plots of the effective EoS with versus $t$ for the different value of parameter $k$. We set $k=3$ (solid line), $k=10$ (dashed line), $k=15$ (dot dashed line) and  $m=1.00672$, $z_{1}=\beta_{0}=-1$ and $z_{2}=1$.} \label{fig:6}
\end{figure}

In this case we choose the form of $f(T,B)$ as a product of power law forms of $B$ and $T$ as $f(T,B)=b_{0}B^kT^n$ where $b_{0}$, $k$ and $n$ are a redefined non-zero constants. Using this form of $f(T,B)$ in Eqs (\ref{nn1})-(\ref{nn5}) we find the following solution
\begin{eqnarray}\label{vec2}
\alpha=-\frac{\beta_{0}}{m+2}Y^{-(m+1)}, \quad \beta=2\beta_{0}Y^{-(m+2)}T,\quad \gamma=\beta_{0}Y^{-(m+2)}B,
\end{eqnarray}
where $\beta_{0}$ is an integration constant and we have a constraint as $n=\frac{1-k}{2}$ ($k\neq1$ and $n\neq0$ which yields a trivial case). Let us try to find some analytical solutions for this case. To do this, we consider three arbitrary functions $z$, $u$ and $w$ depends on the variables of  configuration space as $z=z(Y,T,B)$, $u=u(Y,T,B)$ and $w=w(Y,T,B)$, respectively. Such a transformation allows us to find a cyclic variable so that new Lagrangian can be rewritten in a form such that $L=L(u,w,\dot{z}̇,\dot{u},\dot{w})$. This transformation is always possible if there exist a Noether symmetry. Following this process is described in detail in Ref. \cite{capp2}, one can find the corresponding variables transformation as
\begin{eqnarray}\label{newv}
z=-\frac{Y^{(m+2)}}{(m+2)\beta_{0}}, \quad u=Y^{2(m+2)}T,\quad w=Y^{(m+2)}B,
\end{eqnarray}
where we chose z as a variable cyclic. The original variables are obtained from the Eqs. (\ref{newv}) by converting to the new variables as following
\begin{eqnarray}\label{orw}
Y=\left[-(m+2)\beta_{0}z\right]^{\frac{1}{m+2}}, \quad T=u\left[-(m+2)\beta_{0}z\right]^{-2},\quad B=w\left[-(m+2)\beta_{0}z\right]^{-1}.
\end{eqnarray}
When point-like Lagrangian is rewritten with respect to these new variables, one can obtain it in the following form
\begin{eqnarray}
L=u^{-\frac{k+1}{2}}w^{k-2}\left[\beta_{0}k(m+2)\left(w\dot{u}-2u\dot{w}\right)\dot{z}+\beta_{0}^2(2m+1)w^2\dot{z}^2-\frac{1}{2}uw^2\right]-\rho_{m0}\label{nlag1}.
\end{eqnarray}
We can easily see that the variable $z$ is cyclic in Lagrangian (\ref{nlag1}). This Lagrangian yields the following Euler-Lagrange equations
\begin{eqnarray}\label{new1}
k(m+2)\left(w\dot{u}-2u\dot{w}\right)+2\beta_{0}(2m+1)w^2\dot{z}=\frac{I_{0}}{\beta_{0}}u^{\frac{k+1}{2}}w^{2-k},
\end{eqnarray}
\begin{eqnarray}\label{new2}
(k-1)uw-4\beta_{0}k(m+2)u\ddot{z}-2\beta_{0}^2(k+1)(2m+1)w\dot{z}^2=0,
\end{eqnarray}
\begin{eqnarray}\label{new3}
uw-4\beta_{0}(m+2)u\ddot{z}-2\beta_{0}^2(2m+1)w\dot{z}^2=0,
\end{eqnarray}
\begin{eqnarray}\label{new4}
uw^2+2\beta_{0}k(m+2)\left(w\dot{u}-2u\dot{w}\right)\dot{z}+2\beta_{0}^2(2m+1)w^2\dot{z}^2+2\rho_{m0}u^{\frac{k+1}{2}}w^{2-k}=0,
\end{eqnarray}
here $I_{0}$ is a constant of motion associated with the coordinate $z$. Now, we can rewrite the variables $u$ and $w$ in term of the variable $z$ by using the Eqs. (\ref{toba}) with the condition $X=Y^m$ and Eqs. (\ref{orw}). Then, inserting the results obtained for the $u$ and $w$ into the equations (\ref{new2}) and (\ref{new3}), these equations are identically satisfied. The other equations can be written as
\begin{eqnarray}\label{new5}
2^{\frac{k+1}{2}}\left(-(2m+1)\right)^{\frac{1-k}{2}}\left(m+2\right)^{k}\beta_{0}\dot{z}^{-k}\ddot{z}^{k-2}\left[(k-1)\ddot{z}^{2}-k\dot{z}\dddot{z}\right]=I_{0},
\end{eqnarray}
\begin{eqnarray}\label{new6}
\dot{z}^{2}\ddot{z}^{2}\left[\rho_{m0}\beta_{0}\left(-(2m+1)\dot{z}^{2}\right)^{\frac{k-1}{2}}+2^{\frac{k+1}{2}}k(m+2)^{k}\ddot{z}^{k-2}
\left[\ddot{z}^{2}-\dot{z}\dddot{z}\right]\right]=0.
\end{eqnarray}
For $I_{0}=0$, the non-trivial solution can be easily found from the Eq. (\ref{new5}) as follows
\begin{eqnarray}\label{new7}
z(t)=\frac{z_{2}(t-kz_{1})^{k+1}}{k+1}+z_{3},
\end{eqnarray}
where $z_{i}$ are integration constants and $k\neq-1$. Inserting this solution into the Eq. (\ref{new6}) we can find a constraint $(2m+1)\left[k(m+2)\right]^{k} z_{2}\beta_{0}=\rho_{m0}\left[\frac{-(2m+1)}{2}\right]^{\frac{k+1}{2}}$. By substituting the solution (\ref{new7}) into Eq. (\ref{orw}), we obtain the solution for scale factor on the  $y$ and $z$ axes as
\begin{eqnarray}\label{new9}
Y(t)=\left[-(m+2)\beta_{0}\left(\frac{z_{2}(t-kz_{1})^{k+1}}{k+1}+z_{3}\right)\right]^{\frac{1}{m+2}}.
\end{eqnarray}
On the other hand, the scale factor in the direction of $x$ can be found from the relation $X=Y^m$
\begin{eqnarray}\label{new10}
X(t)=-(m+2)\beta_{0}\left[\frac{z_{2}(t-kz_{1})^{k+1}}{k+1}+z_{3}\right].
\end{eqnarray}
By means of the directional scale factors $X$ and $Y$, the average scale factor for the Universe is defined as $a(t)=(XY^2)^{\frac{1}{3}}=Y^{\frac{m+2}{3}}$ so that we get
\begin{eqnarray}\label{new11}
a(t)=\left[-(m+2)\beta_{0}\left(\frac{z_{2}(t-kz_{1})^{k+1}}{k+1}+z_{3}\right)\right]^{\frac{1}{3}}.
\end{eqnarray}
To analyse the behaviour of the obtained solution, we now examine some cosmological parameters such as directional Hubble, average Hubble parameter, deceleration parameter and equation of state parameter. The directional Hubble parameters $H_{x}=\frac{\dot{X}}{X}$, $H_{y}=H_{z}=\frac{\dot{Y}}{Y}$ and average Hubble parameter $H=\frac{\dot{a}}{a}$ are given by
\begin{eqnarray}\label{new12}
H_{x}=\frac{z_{2}(k+1)(t-kz_{1})^k}{(m+2)\left[z_{2}(t-kz_{1})^{k+1}+(k+1)z_{3}\right]},\quad H_{y}=H_{z}=\frac{H_{x}}{m+2},
\end{eqnarray}
\begin{eqnarray}\label{new13}
H=\frac{z_{2}(k+1)(t-kz_{1})^k}{3\left[z_{2}(t-kz_{1})^{k+1}+(k+1)z_{3}\right]}.
\end{eqnarray}
\begin{figure}
\resizebox{0.45\textwidth}{!}{\includegraphics{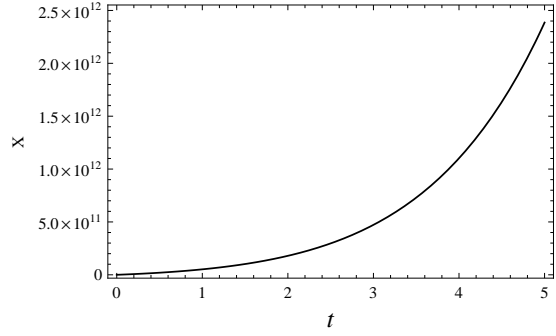}}
\caption{The behaviour of the scale factor in x-direction versus $t$ for the numeric value of parameters $k=10$, $m=1.0672$, $z_{1}=\beta_{0}=-1$ and $z_{2}=1$.} \label{fig:1}
\end{figure}
\begin{figure}
\resizebox{0.45\textwidth}{!}{\includegraphics{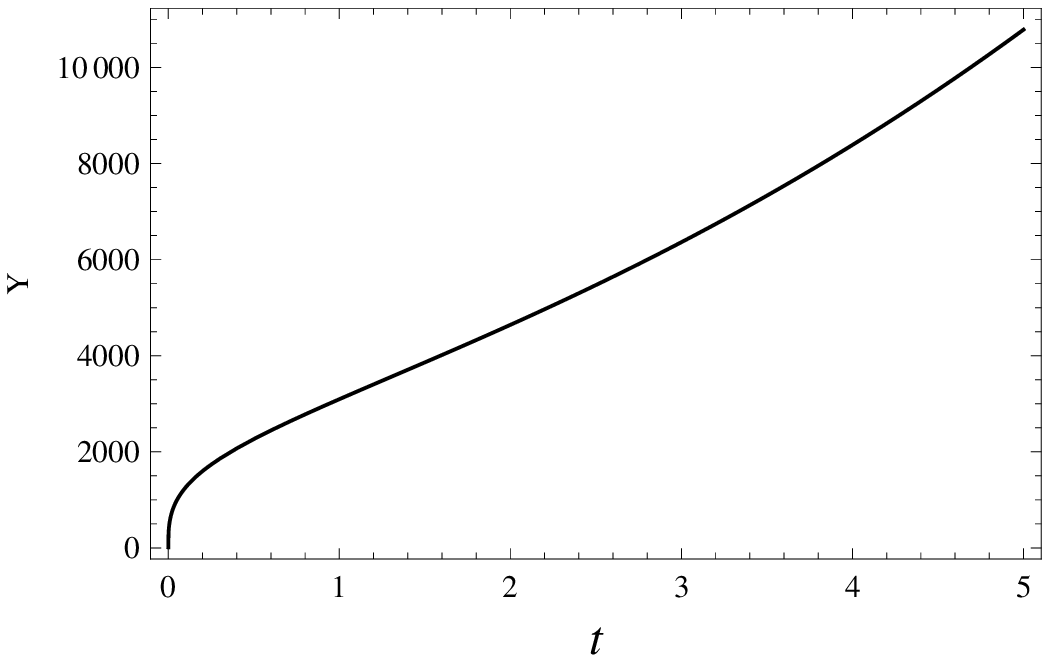}}
\caption{The behavior of the scale factors in $y$ and $z$-direction versus $t$ for the numeric value of parameters $k=10$, $m=1.00672$, $z_{1}=\beta_{0}=-1$ and $z_{2}=1$.} \label{fig:2}
\end{figure}
\begin{figure}
\resizebox{0.45\textwidth}{!}{\includegraphics{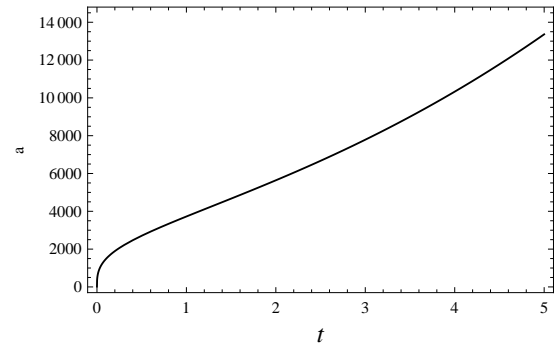}}
\caption{The behavior of the average scale factor versus $t$ by taking $k=10$, $m=1.00672$, $z_{1}=\beta_{0}=-1$ and $z_{2}=1$.} \label{fig:3}
\end{figure}
\begin{figure}
\resizebox{0.45\textwidth}{!}{\includegraphics{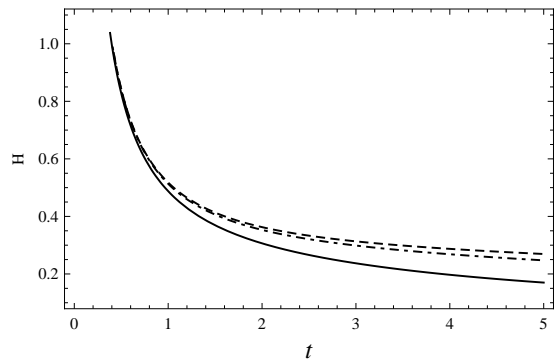}}
\caption{ Evolutions of the average Hubble parameter versus $t$ for the different value of $k$. We set $k=3$ (solid line), $k=10$ (dashed line), $k=15$ (dot dashed line) and  $m=1.00672$, $z_{1}=\beta_{0}=-1$ and $z_{2}=1$.} \label{fig:4}
\end{figure}
The deceleration parameter which is defined by $q =-a \ddot{a}/\dot{a}^2$ plays significant role in describing the nature of the expansion of the universe. The positive value of deceleration parameter indicates a decelerating universe while the negative value shows an accelerating universe. It takes in our model following form
\begin{eqnarray}\label{new14}
q=-1+\frac{3}{k+1}-\frac{3kz_{3}}{z_{2}(t-kz_{1})^{k+1}}.
\end{eqnarray}
The corresponding effective EoS parameter for this model is
\begin{eqnarray}\label{new144}
\omega=-1+\frac{2}{k+1}-\frac{2kz_{3}}{z_{2}(t-kz_{1})^{k+1}}.
\end{eqnarray}
We demonstrate the characteristic behaviour of the present model with respect to cosmic time $t$ via the scale factors along $x$ and $y$ direction and the average scale factor in Figures \ref{fig:1}-\ref{fig:3} by giving some suitable values to the parameters with an initial conditions $a(0)=0$. From these figures we observed that all scale factors increase monotonically when cosmic time increases and approach to infinity as $t\rightarrow\infty$. From Figure \ref{fig:4} which represent the mean Hubble parameter, one can see that it decreases as $t$ increases approaches to zero as $t\rightarrow\infty$. The deceleration parameter $q$ given by Eq. (\ref{new14}), plotted in Figure \ref{fig:5}, tell us that in the early periods of the universe there is a decelerating phase. However, with the time spent it takes values from positive to negative depending on the values of $k$, which shows that our universe has a phase transition at the previous time. On the other hand, the universe enters asymptotically the de Sitter universe for the large values of $k$. We also depict the EOS parameter $\omega$ as a function of the cosmic time for different values of $k$ in Figure \ref{fig:6}. If this parameter is less than $-1/3$, the accelerating expansion of the universe can be generated. Furthermore, the observational constraints show that $\omega$ is around $-1$. When the $\omega$  equal to $-1$, the current universe is defined by the $\Lambda$CDM model where our universe is evolving towards an asymptotically de Sitter future. If $\omega$ lies in the $-1<\omega<-1/3$, the dark energy models are known as  quintessence, but phantom dark energy models have a EoS parameter with $\omega<-1$. As can be seen from the Figure \ref{fig:6}, the effective EoS parameter shows quintessence behaviour of the universe with time and in the late-time limit, it gets close to $\Lambda$CDM model as the value of $k$ increases.
For the special case where the constants $z_{1}=z_{3}=0$, the model also have an important cosmological results. For this specific choice, the average scale factor, deceleration parameter and effective EoS parameter reduce to the following form
\begin{eqnarray}\label{new15}
a(t)=\left(-\frac{(m+2)z_{2}\beta_{0}}{k+1}\right)^{1/3}t^{\frac{k+1}{3}}, \quad q=-1+\frac{3}{k+1} \quad w=-1+\frac{2}{k+1}.
\end{eqnarray}
From above equations, quintessence models of dark energy (i.e. $-1<w<-1/3$) can be achieved for the condition $k>2$ while we have a phantom dark energy models ($w<-1$) for $k>-1$. In these conditions, the universe is both expanding and accelerating. Furthermore, for the interval $-1<k<2$ the model represents decelerating universe.
\subsection{Case 4: $f(T,B)=-T+F(B)$} \label{sec:4.4}
\begin{figure}
\resizebox{0.45\textwidth}{!}{\includegraphics{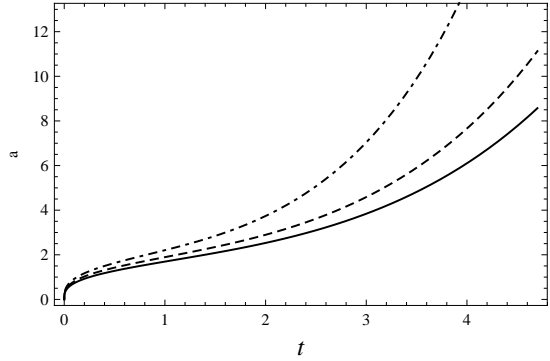}}
\caption{The behaviours of the average scale factor against $t$ for the different value of $m$. We set values $m=1.00672$ (solid line), $m=2.2$ (dashed line) $m=4.5$ (dot dashed line) and $u_{0}=-0.8$, $a_{0}=I_{0}=u_{1}=1$.} \label{fig:7}
\end{figure}

\begin{figure}
\resizebox{0.45\textwidth}{!}{\includegraphics{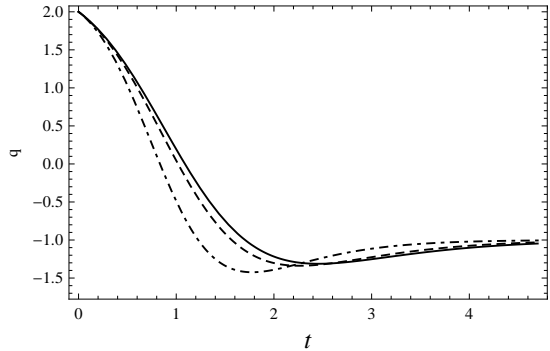}}
\caption{The behaviours of the deceleration parameter versus $t$ for the different value of $m$ by taking values $m=1.00672$ (solid line), $m=2.2$ (dashed line) $m=4.5$ (dot dashed line) and $u_{0}=-0.8$, $a_{0}=I_{0}=u_{1}=1$.} \label{fig:8}
\end{figure}

\begin{figure}
\resizebox{0.45\textwidth}{!}{\includegraphics{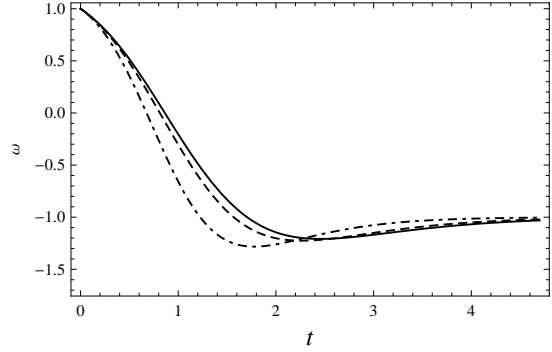}}
\caption{The behaviours of the effective EoS parameter versus $t$ for the different value of $m$. We set values $m=1.00672$ (solid line), $m=2.2$ (dashed line) $m=4.5$ (dot dashed line) and $u_{0}=-0.8$, $a_{0}=I_{0}=u_{1}=1$.} \label{fig:9}
\end{figure}

Finally, we consider an interesting model that includes the torsion scalar plus a function of the boundary term. If $F(B)$ is a linear in $B$, then the model reduce to the standard general relativity theory. By placing this model into the Noether symmetry equations we conclude that the vector field (\ref{vec}) does not comprise its component $\beta$. Thus, Noether symmetry condition (\ref{nn1})-(\ref{nn5}) generate the following solutions for the vector field and the function $F(B)$
\begin{eqnarray}\label{new16}
\alpha=a_{0}Y^{-(m+1)}, \quad \gamma=-a_{0}(m+2)Y^{-(m+2)}B,
\end{eqnarray}
\begin{eqnarray}\label{new17}
F(B)=b_{0}B+\frac{(2m+1)Bln(B)}{(m+2)^2},
\end{eqnarray}
where $a_{0}$ and $b_{0}$ are an integration constants. Considering the above solutions (\ref{new16}) allows us to do the following coordinate transformations
\begin{eqnarray}\label{new18}
z=\frac{Y^{m+2}}{a_{0}(m+2)}, \quad u=Y^{m+2}B.
\end{eqnarray}
So the Lagrangian in the transformed variables for the present model takes the suitable form
\begin{eqnarray}\label{new19}
L=\frac{(2m+1)\left(2a_{0}\dot{u}\dot{z}-u^2\right)}{(m+2)^2u}-\rho_{m0},
\end{eqnarray}
in which $z$ is cyclic variable. The Euler-Lagrange equations relative to the Lagrangian (\ref{new19}) are
\begin{eqnarray}\label{new20}
\frac{2(2m+1)a_{0}\dot{u}}{(m+2)^2u}=I_{0},
\end{eqnarray}
\begin{eqnarray}\label{new21}
2a_{0}\ddot{z}+u=0,
\end{eqnarray}
\begin{eqnarray}\label{new22}
(2m+1)\left(2a_{0}\dot{u}\dot{z}+u^2\right)+(m+2)^2\rho_{m0}u=0,
\end{eqnarray}
where $I_{0}$ is a constant of motion for the present model. The general solution of the Eqs. (\ref{new20})-(\ref{new22}) is
\begin{eqnarray}\label{new23}
u(t)=u_{0}e^{st}, \quad z(t)=-\frac{u_{0}e^{st}}{2s^2a_{0}}+u_{1}t+u_{2},
\end{eqnarray}
with the constrain $\rho_{m0}+I_{0}u_{1}=0$. Here, $u_{i}$ are integration constants and we define $s=\frac{I_{0}(m+2)}{2a_{0}(2m+1)}$. Going back to physical variables,one can find the solution in the following form
\begin{eqnarray}\label{new24}
Y(t)=\left[a_{0}(m+2)\left(-\frac{u_{0}e^{st}}{2s^2a_{0}}+u_{1}t+u_{2}\right)\right]^{\frac{1}{m+2}}.
\end{eqnarray}
The average scale factor is
\begin{eqnarray}\label{new25}
a(t)=\left[a_{0}(m+2)\left(-\frac{u_{0}e^{st}}{2s^2a_{0}}+u_{1}t+u_{2}\right)\right]^{\frac{1}{3}}.
\end{eqnarray}
For this model, we obtain the deceleration parameter
\begin{eqnarray}\label{new26}
q=-1+\frac{6sa_{0}\left[u_{0}e^{st}\left(u_{1}st+u_{2}s-2u_{1}\right)+2sa_{0}u_{1}^2\right]}{\left(u_{0}e^{st}+2sa_{0}u_{1}\right)^2},
\end{eqnarray}
and the effective EoS parameter
\begin{eqnarray}\label{new27}
\omega=-1+\frac{4sa_{0}\left[u_{0}e^{st}\left(u_{1}st+u_{2}s-2u_{1}\right)+2sa_{0}u_{1}^2\right]}{\left(u_{0}e^{st}+2sa_{0}u_{1}\right)^2}.
\end{eqnarray}
Similarly to the behaviour of cosmological solutions in the previous model, the average scale factor with an initial conditions $a(0)=0$, shown in Figure \ref{fig:7}  is a monotonically increasing function of time. Evolution of the deceleration parameter as a function of time for different values of anisotropy parameter $m$ depicted in Figure \ref{fig:8}. It can be seen from this figure that our model shows the transition of $q$ from the decelerating to the accelerating phase and in the limit $t\rightarrow\infty$ its evolution becomes de Sitter Universe. Figure \ref{fig:9} shows behaviour of the effective EoS parameter with respect to cosmic time $t$ for the different values of $m$. From this figure, we observe that crossing of the phantom divide line $\omega=-1$ can be addressed in this model described by the Noether symmetry solution.

\section{Summary and Conclusion}\label{sc5}
The modified theories of gravity that are constructed to describe the accelerated expansion of the universe are of great importance. One of these theories is the new generalization of teleparallel gravity including both functions of the torsion scalar and the boundary term in the form of $f(T,B)$ introduced by Bahamendo et al. \cite{baha1}. In this work, we considered the cosmology constructed from $f(T,B)$ theory of gravity with anisotropy background. For this purpose, we considered LRS Bianchi type I cosmological model in the presence of matter dominant universe and due to highly non-linear and complicated field equations, we used a physical assumption $X=Y^m$. The Noether symmetry approach is well known to be an important method for solving dynamical equations. Here, we discussed the Noether symmetry equations for two interesting cases of the $f(T,B)$ gravity theory. The first case is to $f(T,B)=b_{0}B^kT^{\frac{1-k}{2}}$ where $b_{0}$ and $k$ are an arbitrary real number. By introducing cyclic variables, we obtained some exact cosmological solutions of the corresponding field equations using the Noether symmetry approach. The second interesting case we are interested in is the form $f(T,B)=-T+F(B)$ in which $F(B)$ is only the function of $B$. We determined the explicit form of $F(B)$ and solved the field equations via Noether symmetry method. We also presented some cosmological parameters for the two cases and depicted the graphical behaviours of the models. The main and interesting feature of these solutions is that they describe an accelerating expansion of the universe. We also stress that phantom divide crossing can be realised in the second case but it is not crossed in the first case.

\section*{Acknowledgements}
I am grateful to Dr. Timur Sahin for fruitful discussions. This work was supported by the scientific research projects units of Akdeniz University.

\vspace{6pt}





\end{document}